\documentclass[10pt,emptycopyrightspace]{ewsn-proc}

\usepackage{graphicx}
\usepackage{balance}
\usepackage{comment}
\usepackage{xcolor}
\usepackage{amssymb}
\usepackage{booktabs} 
\usepackage{subfigure,wrapfig}
\usepackage[hyphens]{url}
\usepackage{amsmath}

\newcommand*{\revise}{\textcolor{black}}
\definecolor{LightCyan}{rgb}{0.0,1,1}
\definecolor{LightGray}{rgb}{0.8,0.8,0.8}

\numberofauthors{2}
\author{
\alignauthor Dali Ismail \\
    \affaddr{Department of Computer Science\\Wayne State University\\Detroit, Michigan, USA }\\
\alignauthor Abusayeed Saifullah \\
    \affaddr{Department of Computer Science\\Wayne State University\\Detroit, Michigan, USA }\\
}

\title{Handling Mobility in Low-Power Wide-Area Network}

\begin{document}

\maketitle

\begin{abstract}
Despite the proliferation of mobile devices in various wide-area Internet of Things applications (e.g., smart city, smart farming), current Low-Power Wide-Area Networks (LPWANs) are not designed to effectively support mobile nodes. In this paper, we propose to handle mobility in SNOW (Sensor Network Over White spaces), an LPWAN that operates in the TV white spaces. 
SNOW supports massive concurrent communication between a base station (BS)  and numerous low-power nodes through a distributed implementation of OFDM.  In SNOW, inter-carrier interference (ICI) is more pronounced under mobility due to its OFDM based design. Geospatial variation of white spaces also raises challenges in both intra- and inter-network mobility as the low-power nodes are not equipped to determine white spaces. To handle mobility impacts on ICI, we propose a dynamic carrier frequency offset estimation and compensation technique which takes into account Doppler shifts without requiring to know the speed of the nodes. We also propose to circumvent the mobility impacts on geospatial variation of white space through a mobility-aware spectrum assignment to nodes.  To enable mobility  of the nodes across different SNOWs, we propose an efficient handoff management through a fast and energy-efficient BS discovery and quick association with the BS by combining time and frequency domain energy-sensing. Experiments through SNOW deployments in a large metropolitan city and indoors show that our proposed approaches  enable mobility across multiple different SNOWs and provide robustness in terms of reliability, latency, and energy consumption under mobility.

\end{abstract}

%
%
%

\section{Introduction}\label{sec:introduction}
Low-Power Wide-Area Network (LPWAN) is an enabling technology for wide-area  Internet-of-Things (IoT) applications such as smart city, agricultural IoT, and industrial IoT offering long-range (several miles),  low-power, and low-cost communication~\cite{ismail2018low}. With the fast growth of IoT, multiple  LPWAN technologies have emerged recently such as LoRa~\cite{lorawan,icnp20}, SigFox~\cite{sigfox}, IQRF~\cite{iqrf}, RPMA~\cite{rpma}, DASH7~\cite{dash7}, Weightless-N/P~\cite{weightless}, Telensa~\cite{telensa} in the ISM band,  and  EC-GSM-IoT~\cite{ecgsmiot}, NB-IoT~\cite{nbiot},  and LTE Cat M1~\cite{cat, LTE_advancedpro} in the licensed cellular band. To avoid the {\slshape crowd} in the {\slshape limited} ISM band and the {\slshape cost} of licensed band, {\slshape  SNOW (Sensor Network Over White spaces)} is an LPWAN architecture  to support scalable wide-area IoT  over the TV white spaces \cite{ton_snow, snow, snow2}.  {\slshape White spaces} are the allocated but locally unused TV spectrum (54 - 698 MHz in the US) \cite{FCC_first_order, fcc_second_order}. They usually have wide and less crowded spectrum in rural and most urban areas, with an abundance in rural areas \cite{ws_sigcomm09}.

With a wide range of supported applications, IoT is integrating more mobile nodes/devices in different domains (e.g. agriculture~\cite{agriculture,farmbeats}, connected vehicle~\cite{connectedvehicle}, healthcare~\cite{health}, smart city~\cite{smart_city}). For example, in agricultural IoT, the use of drones and tractors is rapidly increasing~\cite{monsanto,farmbeat,drone1,farmbeats}. It is expected that by the year 2050, there will be more than 3 billion wearable sensors \cite{mobilelpwan1}. The cellular-based LPWANs rely on wired infrastructure to handle mobility. Such infrastructure is often not available in rural and remote areas (e.g., farms, oil fields, etc.). In others,  mobility introduces challenges that are not well-addressed yet. Study on LoRa shows that its performance is susceptible even to minor human mobility~\cite{mobilelpwan1, mobilelpwan2}.

In this paper, \revise{we propose to handle mobility in LPWAN in the white spaces (in the US) considering SNOW}. With the rapid growth of IoT,  LPWANs will suffer from crowded spectrum due to long range, making it critical to exploit white spaces. SNOW is a highly scalable LPWAN over the white spaces which enables massive concurrent communication between a base station (BS)  and numerous low-power nodes. It is available as an open-course implementation \cite{opensource}.  Its physical layer is designed based on a Distributed implementation of OFDM (orthogonal frequency division multiplexing)  for multi-user access, called D-OFDM. The BS operates on a wide band spectrum which is split into many orthogonal narrowband subcarriers. A node (non-BS) transmits and receives on a subcarrier.  In SNOW, inter-carrier interference (ICI) is more pronounced under mobility due to its OFDM based design. Geospatial variation of white spaces also raises challenges in both intra- and inter-network mobility as the low-power nodes are not equipped to determine white space. For example, to enable mobility across different SNOWs, it is challenging for a node to scan the wide spectrum of the TV band to discover a new BS. Besides, different BSs may be using different subcarrier widths, which may result in subcarrier misalignment between the mobile node and the new BS.

In this paper, we address the challenges mentioned above to handle mobility in SNOW. Specifically, we make the following new contributions. 

\begin{itemize}

\item To handle mobility impacts on ICI, we propose a dynamic CFO (Carrier Frequency Offset ) estimation and compensation technique for SNOW which takes into account Doppler shifts under non-uniform speeds without requiring to know the speed of the nodes. To circumvent the mobility impacts on geospatial variation of white space within the same SNOW, we propose a mobility-aware subcarrier assignment to the nodes.

\item To handle inter-SNOW (inter-network)  mobility, we propose an energy-efficient and fast BS discovery technique that considers the trade-off between discovery latency and energy consumption to allow efficient handoff management. Our approach utilizes the spectrum information by combining the received signal features to distinguish between primary users (TV stations) and a SNOW BS. 
We also propose a lightweight cross-layer technique feasible at the energy-constrained SNOW nodes to handle subcarrier alignment by combining time and frequency domain energy-sensing.

\item We implement our proposed mobility handling techniques on SNOW devices and  perform experiments by deploying SNOW in two environments - a large metropolitan city and an indoor testbed. The experimental results show that our approaches enable mobility across multiple different SNOWs. The results also show an improvement of reliability from  80\% to 96.6\% when our dynamic CFO estimation and compensation is incorporated.

\end{itemize}

The rest of the paper is organized as follows. Section~\ref{sec:related} overviews related work. Section~\ref{sec:background} presents an overview of SNOW. Section~\ref{sec:system} describes the system model. Section~\ref{sec:approach} presents our mobility approach. Section~\ref{sec:experiments} presents the experiments. Section~\ref{sec:conclude} concludes the paper.

\section{Related Work}\label{sec:related}

Many studies focused on handling mobility in Wireless Sensor Networks (WSNs)~\cite{msmac, mmac, tdmamac, smac, mcmac, disco} (more can be found in survey~\cite{mobility_survey2, mobility_survey}) and ad hoc network ~\cite{Camp02asurvey, vadhoc_survey}. In WSN or WiFi  networks, a client has to scan only a limited/fixed number of channels to discover a new BS. However, those approaches are not directly applicable to LPWAN. To handle mobility across networks, cellular LPWANs rely on wired infrastructure. Non-cellular LPWANs are not yet handling it well. Real experiments show that their performance is susceptible even to minor human mobility~\cite{mobilelpwan1}. Spectrum mobility studied  in~\cite{cmcs, prospect} for cognitive networks enables secondary users to change the operating frequencies, and is different from device mobility. Device mobility was studied in~\cite{database2} for white space network, where every device primarily relies on the database to determine the white spaces. A mobile device adds a protection range of $\delta d$ so that any channel blocked within distance $\delta d$ of current location is not used. Note that this approach does not work for SNOW as the nodes have no direct access to the Internet (and database). It first has to discover a BS, associate with it, and rely on it for spectrum access. Additionally, there has been much work on  channel rendezvous in cognitive radio~\cite{rendezvous1, rendezvous2, rendezvous3, gameapproach}. Due to technology-specific nature of SNOW, these techniques cannot be applied to a SNOW.

\emph{Senseless}~\cite{senseless} is an infrastructure based white space network system where the devices do not rely on sensing to determine the availability of white space. They use geo-location service to calculate white space availability at any location. \emph{Senseless} then disseminates the availability information to each device in the network. To address the mobility challenges in white space, Senseless suggests that every device adds a protection range to determine the availability of white spaces while it is mobile. This could lead to a huge spectrum waste depending on the size of the protection area. SNOW differ from Senseless in that it considers infrastructure-less network system where BSs are not connected. Besides, the D-OFDM based design requires a different approach for mobility in  SNOW. Also, we incorporate an energy-efficient sensing approach along with the geo-location service to efficiently handle nodes mobility. To date, inter-network mobility for non-cellular LPWAN remains mostly unexplored and it was never studied for SNOW.

\section{A Brief Overview of SNOW}\label{sec:background}

 \begin{figure}
\centering
\includegraphics[width=0.39\textwidth]{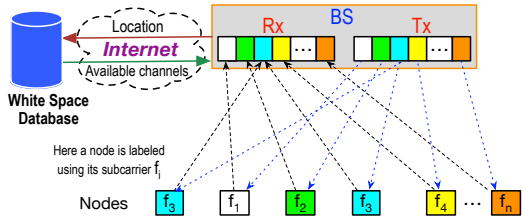}
\caption{The SNOW architecture.}
\label{fig:arch}
\end{figure}

 \begin{figure}
\centering
\includegraphics[width=0.29\textwidth]{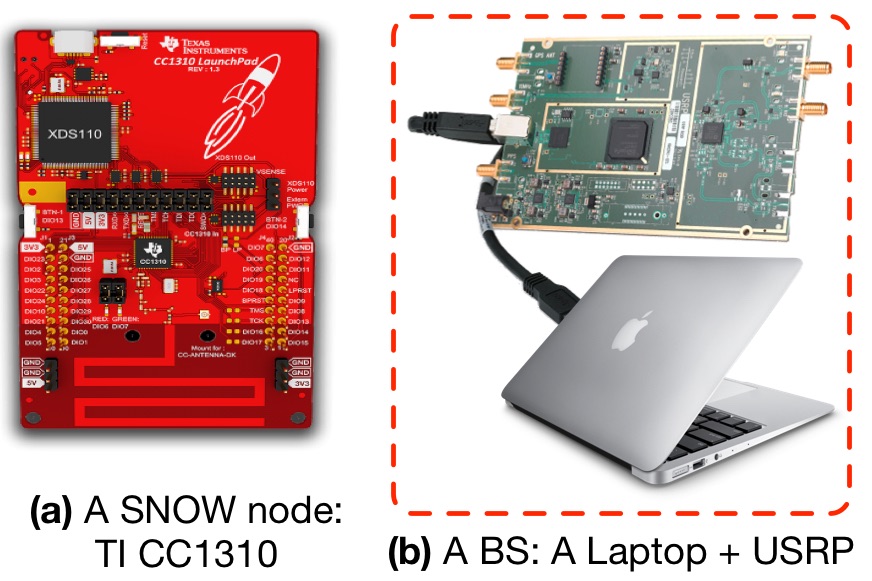}
\caption{SNOW hardware.}
\label{fig:hardware}
\end{figure}

 Here we provide a brief overview of the SNOW architecture \cite{snow, snow2, ton_snow}. Due to long transmission (Tx) range (several miles at 0dBm), the nodes in SNOW are directly connected to the BS, forming a star topology as shown in Fig.~\ref{fig:arch}. We use {\bf `node'} to indicate a sensor node.  The BS periodically determines white spaces by providing locations of its own and of all other nodes in a cloud-hosted database through the Internet. It uses wide white space spectrum as a single wide channel that is split into narrowband orthogonal subcarriers, each of equal spectrum width (bandwidth). Each node has a single half-duplex narrowband radio. It sends/receives on a subcarrier. The nodes are power-constrained, and do not do spectrum sensing or cloud access. As shown in Fig. \ref{fig:arch}, the BS uses two radios operating on the same spectrum --  one for only transmission (called {\slshape Tx radio})  and the other for only reception (called {\slshape Rx radio}) -- to facilitate concurrent bidirectional communication.

The physical layer (PHY) of SNOW is designed based on a {\bf D}istributed implementation of {\bf OFDM} for multi-user access, called {\bf D-OFDM}. D-OFDM  splits a wide spectrum into numerous narrowband orthogonal subcarriers enabling parallel data streams to/from numerous distributed nodes from/to the BS.  A subcarrier bandwidth is in kHz (e.g., 50kHz, 100kHz, 200kHz, or so depending on packet size and needed bit rate).  The nodes transmit/receive on orthogonal subcarriers, each using one. A subcarrier is modulated using Binary Phase Shift Keying (BPSK) or Amplitude Shift Keying (ASK).  If the BS spectrum is split into  $m$  subcarriers,  it can receive from $m$ nodes simultaneously using a single antenna. Similarly, it can transmit different data on different subcarriers through a single transmission. Currently, the sensor nodes in SNOW use a very simple and lightweight CSMA/CA based MAC (media access control) protocol like the one used in TinyOS~\cite{tinyos}. 

SNOW was implemented on two hardware platforms \cite{iotdi2019} -- USRP (universal software radio peripheral) \cite{usrp} using GNU radio \cite{gnuradio} and TI CC1310~\cite{cc1310} (Figure \ref{fig:hardware}). A dual-radio USRP connected to Raspberry PI or Laptop is used as the BS.  A CC1310 device or a single-radio USRP can be used as a SNOW node. CC1310 is a tiny, cheap ($<$\$30), and commercially off-the-shelf (COTS) device with a programmable PHY. We have adopted the open-source implementation of SNOW that is available at \cite{opensource}.

\section{System Model}\label{sec:system}
\begin{figure}[!htb]
\centering
\includegraphics[width=0.4\textwidth]{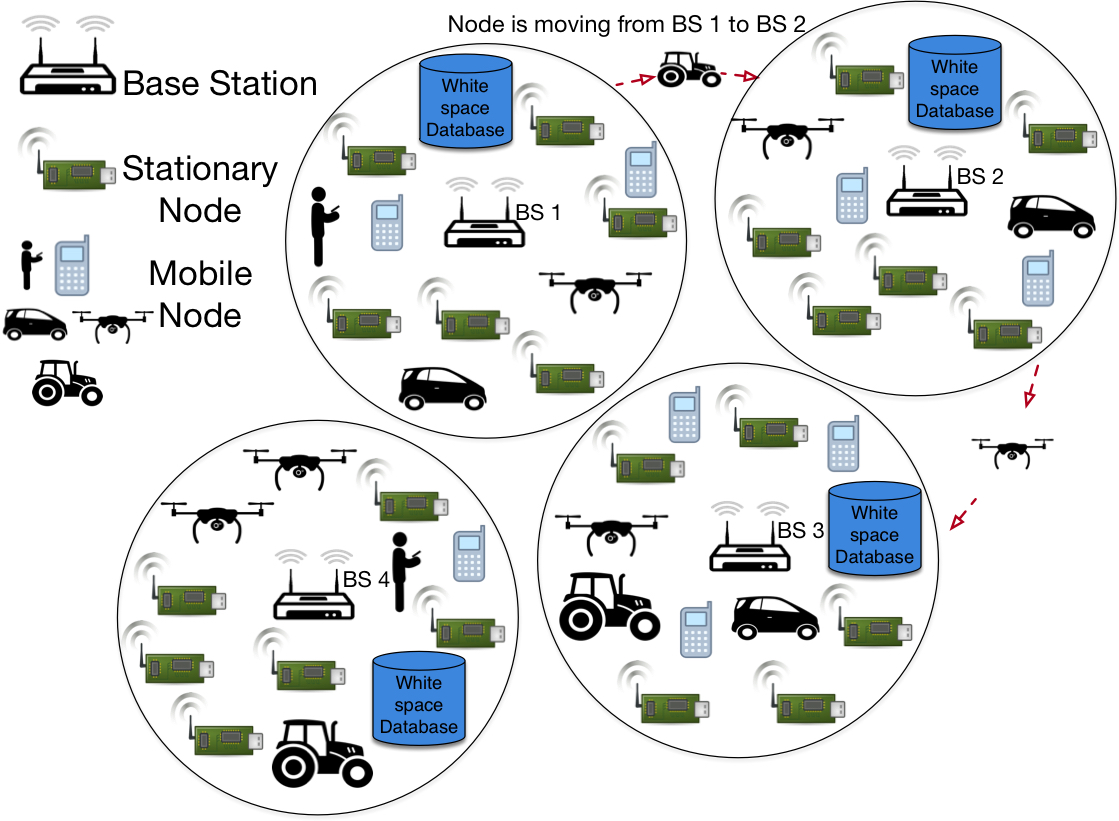}
\caption{Inter-SNOW mobility: the figure shows multiple SNOWs where a mobile node is moving from one SNOW to another. }
\label{fig:snow}
\end{figure}

We consider multiple independent and uncoordinated SNOWs. Each SNOW is having its own BS and associated nodes. The nodes are battery-powered and thus have energy constraints. We assume the existence of both mobile and stationary nodes in SNOW. A mobile node can move from one SNOW to any SNOW, as depicted in Figure~\ref{fig:snow}. Since the BS has a long-range, it can cover a wide area. Hence, we assume the BSs are stationary. Each node is equipped with a half-duplex white space radio. The BS and its associated nodes form a star topology where nodes can directly communicate with the BS. 

The BSs are independent, connected to the Internet, and directly connected to a power source. Each BS uses a wide channel. This channel is split into narrowband channels/subcarriers of equal width. Each node is assigned a single subcarrier for transmission and reception to/from the BS. The nodes are kept simple by offloading the complexities to the BS. The BS determines the availability of white space at its location by querying a cloud-hosted database through the Internet. We assume each BS knows the location of its associated nodes either manually or through existing localization techniques~\cite{wsnlocalizationsurvey}. However, we are not considering localization in this paper.

\section{Handling Mobility}\label{sec:approach} 
In this section, we present our techniques to address mobility challenges in both intra- and inter-SNOW mobility. We propose to address those through lightweight cross-layer approaches (MAC-PHY design) feasible at energy-constrained nodes. First, we present our mobility handling within the same SNOW (intra-SNOW mobility), and then we will present mobility handling across SNOWs (inter-SNOW mobility).

\subsection{Handling Mobility within the Same SNOW}
Mobility affects communication reliability even when a node moves within the same network due to ICI occurred in OFDM subcarriers and also due to geospatial variation of spectrum within the same network. We address both scenarios as described below.

\subsubsection{Handling Mobility Impacts on ICI}
\begin{figure}[!htb]
\centering
\includegraphics[width=0.28\textwidth]{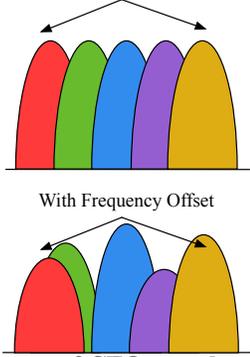}
\vspace{-0.05in}
\caption{The impact of CFO on subcarriers orthogonality.}
\label{fig:cfo}
\end{figure}

ICI is introduced mainly due to the CFO, which stems from the frequency mismatch between the transmitter and receiver oscillators due to hardware imperfections and the {\slshape\bf Doppler shift} which is a function of their relative speed. In SNOW, the subcarriers loose their orthogonality due to such CFO as shown Figure~\ref{fig:cfo}. Hence, to improve an OFDM system's performance, CFO needs to be estimated and compensated. Currently, in SNOW, CFO estimation and compensation is done considering stationary nodes or assuming node speeds are known  \cite{iotdi2019}. However, SNOW nodes are energy-constrained and low-cost and may not be equipped to determine speeds.  We first give an overview of the adopted CFO estimation technique and then describe our Doppler shift handling approach without the need to know the speed of the nodes or under non-uniform speed.

SNOW uses training symbols (preamble) for CFO estimation. Due to its distributed and asynchronous nature, CFO estimation in D-OFDM is done slightly differently than traditional OFDM. CFO estimation in D-OFDM is done when a node joins the network. SNOW uses one (or more) subcarrier for a node joining the network, called {\slshape join subcarrier}, that does not overlap with any other subcarrier. Each node joins the network by first communicating with the BS on a join subcarrier. Each way, communication (BS to node and node to BS) follows a preamble used to estimate CFO on join subcarrier. Specifically, preamble from a node to BS allows to estimate CFO at the BS, and that from BS to a node allows to estimate CFO at the node on the join subcarrier. Later, based on the CFO on a join subcarrier,  the CFO on a node's assigned subcarrier is determined. CFO estimation technique for both upward and downward communication is similar. However,  CFO compensation approaches in upward and downward communication are different (refer to~\cite{iotdi2019} for detailed explanation).

First, we explain how CFO is estimated on a join subcarrier $f$. Since it does not overlap with other subcarriers, it is ICI-free. If $f_{Tx}$ and $f_{Rx}$ are the frequencies at the transmitter and at the receiver, respectively, then their frequency offset $\Delta f = f_{Tx} - f_{Rx}$. For transmitted signal $x(t)$, the received signal $y(t)$ that experiences a CFO of $\Delta f$ is given by
\begin{equation}\label{eqn:eqn1}
y(t) = x(t)e^{j2\pi\Delta ft}
\end{equation}

 $\Delta f$ is estimated based on short and long preamble approach using time-domain samples.  A 32-bit preamble is divided into two equal parts, each of 16 bits. First part is for coarse estimation and the second part is for finer estimation of CFO~\cite{frequency}. Considering $\delta t$ as the short preamble duration,
$$ y(t-\delta t)  = x(t) e^{j2\pi \Delta f (t-\delta t)}.$$ 
Since $y(t)$ and $y(t-\delta t)$ are known at the receiver, 
\begin{align*}
y(t-\delta t) y^*(t)  & = x(t) e^{j2\pi \Delta f (t-\delta t)}       x^*(t) e^{-j2\pi  \Delta f t}\\
                            & = |x(t)|^2  e^{j 2\pi  \Delta f -\delta t }
\end{align*}
Taking angle of both sides, 
$$\sphericalangle  y(t-\delta t) y^*(t)   =  \sphericalangle     |x(t)|^2  e^{j 2\pi  \Delta f -\delta t }  =      - 2\pi  \Delta f \delta t.        $$
$$\text{Thus,~~~~~~~~~~~~~~~~~~~~~~~}  \Delta f   =  - \frac{\sphericalangle  y(t-\delta t) y^*(t) }{2\pi\delta t}~~~~~~~~~~~~~~~~~~~~~~~~~~~~~~  $$

A SNOW node calculates the CFO on join subcarrier $f$ using the preambles from the BS to the node using the above approach. In upward communication, the time-domain samples are used for CFO estimation on the join subcarrier $f$ at the BS based on the above approach. Then the {\slshape ppm (parts per million)} on the receiver's (BS or SNOW node) crystal is given by 
$ \text{ppm} = 10^6  \frac{\Delta f}{f} $. Thus, the receiver (BS or a node) calculates $ \Delta f_i$ on subcarrier $f_i$ as  
$$ \Delta f_i =  \frac{f_i * \text{ppm}}{10^6}. $$
Thus the BS and a SNOW node that is assigned subcarrier $f_i$ calculates CFO on $f_i$ on its respective side. 
As the nodes asynchronously transmit to the BS, doing the CFO compensation for each subcarrier at the BS is quite tricky. Hence, a simple feedback approach for proactive CFO correction in upward communication is adopted. In this approach, a transmitting node adjusts its frequency based on $\Delta f_i$ when transmitting on subcarrier $f_i$  so that the BS does not have to compensate for $\Delta f_i$.

Since mobility causes Doppler shift in frequency contributing further to CFO, CFO has to be estimated using the above approach while a node moves. If a node moves at speed $v$, such 
Doppler Frequency Offset (DFO), denoted by $\delta_f (v)$, is upper-bounded  by 
\begin{equation}\label{eqn:doppler}
\delta_f (v) = \dfrac{v}{c} f_{c}
\end{equation}
where, $c$ is the speed of light, and $f_{c}$ is the carrier frequency. Therefore, considering  $\Delta f_i$ as the CFO when a node is stationary, it experiences a total CFO of   $\Delta f_i  +  \delta_{f_i} (v) $  when it moves at speed $v$. Therefore, to account for this total CFO, the node needs to know its speed. Besides, when the speed changes, $ \delta_{f_i}$ has to be recalculated. But, being energy-constrained and low-cost, SNOW nodes are not equipped to determine their speeds. Hence, we rely on the observation that when CFO is estimated for a moving node using the above CFO estimation technique, its estimation includes both $\Delta f_i $  and $\delta_{f_i}$, resulting in a CFO of $\Delta f_i  +  \delta_{f_i}$. Thus, the node does not need to know its speed. If the node's speed changes, then the total CFO changes and we need re-estimate. However, the node has no way to determine if its speed increases or decreases. To handle this challenge, we enable each node to periodically estimate the CFO. \revise{This period can be set as a tunable system parameter that can be adjusted dynamically}. Estimating CFO periodically will ensure that if the speed changes, the new CFO calculation takes the new speed into account.

\begin{figure*}[!htb]
  \centering
    \subfigure[\small Subcarrier aligned with BS\label{fig:align1}]{
    \includegraphics[width=0.33\textwidth]{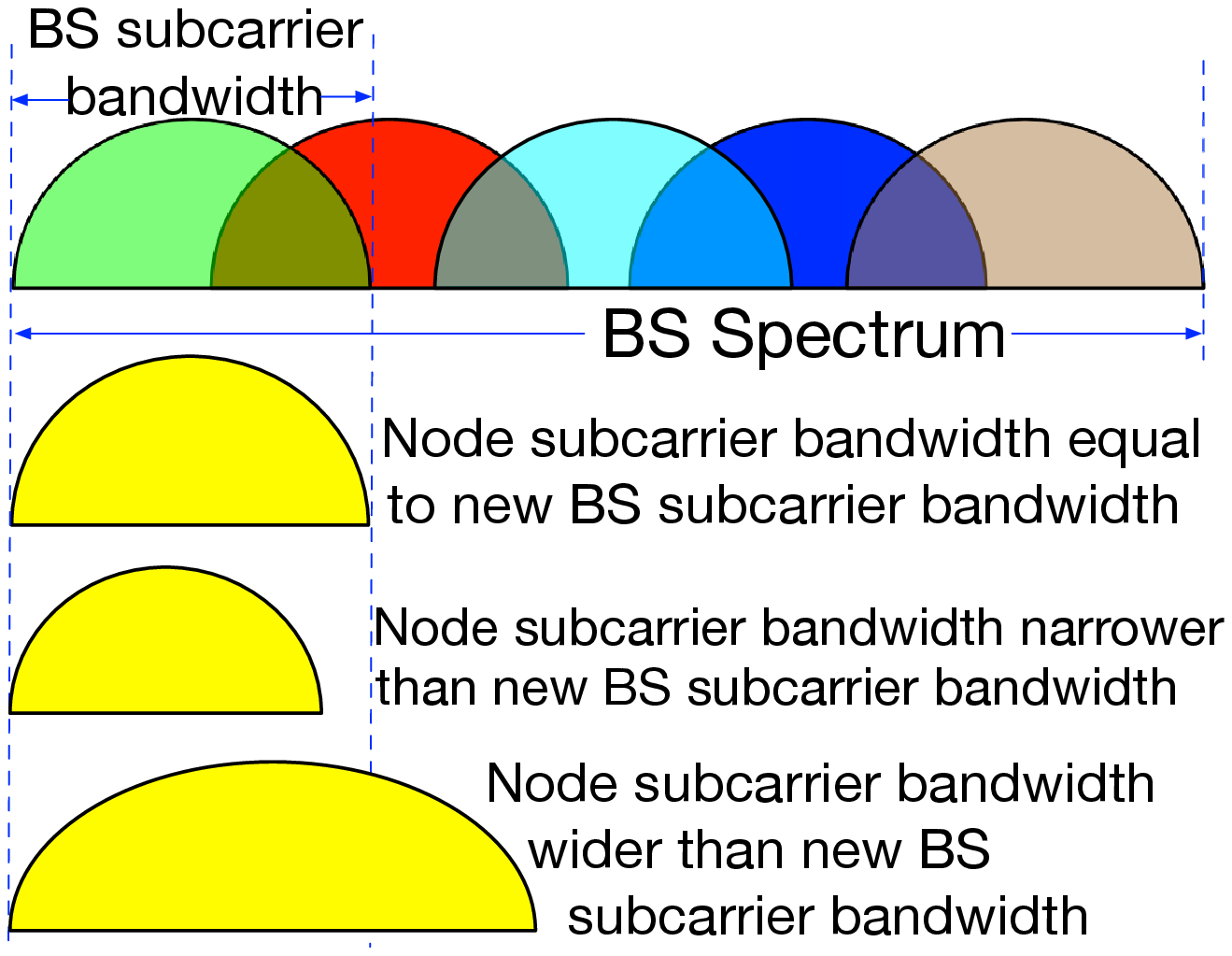}
      }
    \hfill
      \subfigure[\small Scenarios showing unalignment\label{fig:align2}]{
        \includegraphics[width=.33\textwidth]{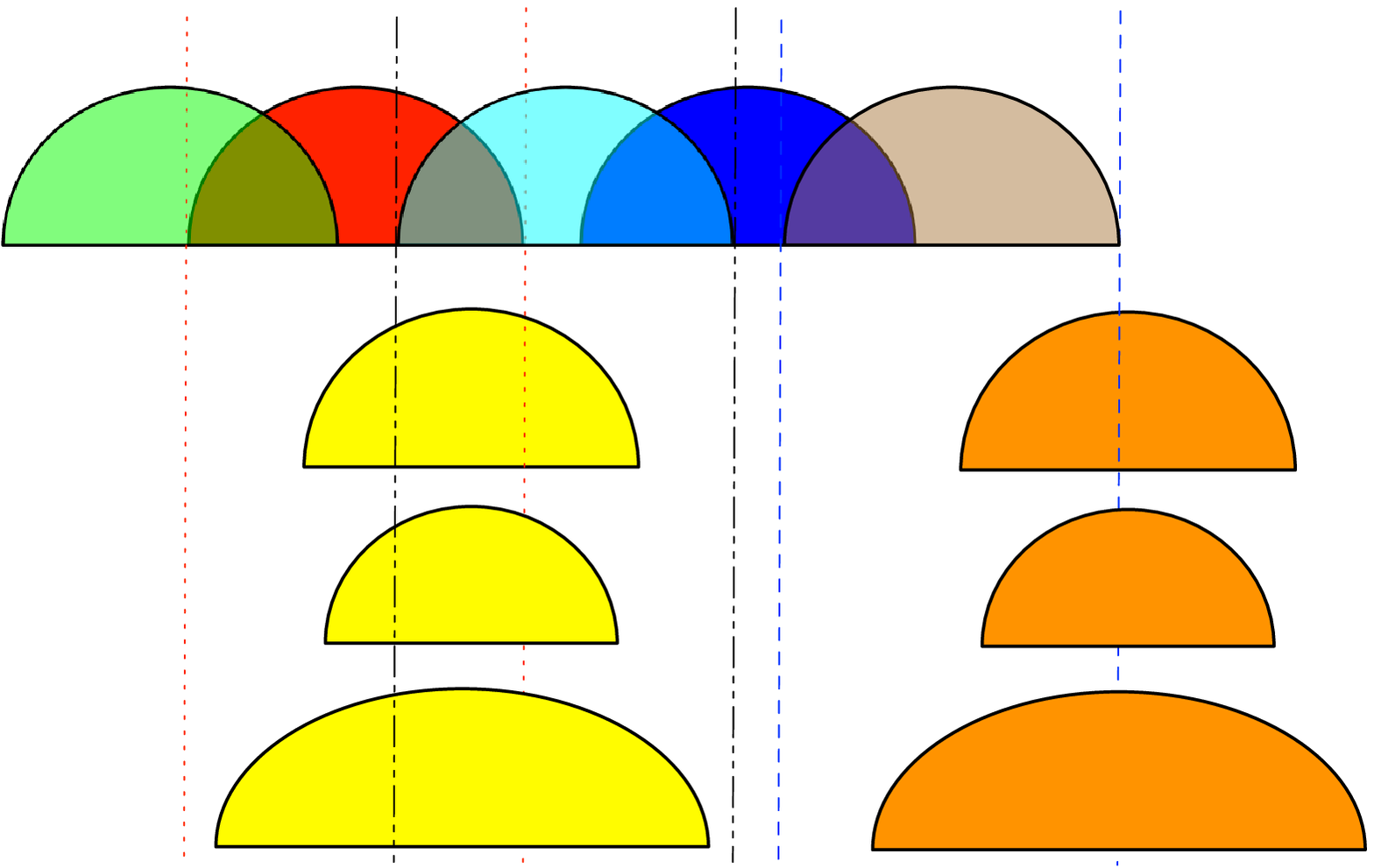}
      }
    \caption{Alignment between the node subcarrier and BS subcarrier.}
    \label{fig:align}
 \end{figure*}

\subsubsection{Handling Mobility Impacts on Geospatial Variation of White Spaces} 

Due to long range, a node's mobility even within the same network affects the spectrum availability. For example, a subcarrier that is assigned to a node at a particular place may not be available if the node moves to another location within the range of the same BS (i.e., within the same SNOW). Currently, the BS assigns subcarriers to the nodes without considering their mobility.  This may affect the communications of the mobile nodes. Namely, if a node is highly mobile and may move  anywhere inside the network but is assigned a subcarrier which is available only in a few locations, its subcarrier assignment is not much useful. To handle this problem  due to geospatial variation of white spaces, we propose a mobility-aware subcarrier assignment policy as follows.

Note that the BS is already assumed to know the location information of its coverage area. We also assume that the BS knows the degree or rate of mobility of each node  (i.e, how much mobile the node is). A node can provide a rough estimate of its mobility when it joins the network. \revise{For example, in agricultural IoT, the system designer knows the number of mobile nodes (e.g., tractors and drones) and each node's mobility rate in a specific geographical area.} The BS orders the nodes based on their mobility, where the stationary nodes come first and the most mobile node is the last. The BS then orders the subcarriers based on their availability, from the least widely available (inside its communication range) subcarrier to the most widely available one. That is, the subcarrier that is available in the minimum number of locations comes first and that available in the maximum number of locations (inside the network) comes at the last of this order. If there are $m$ subcarriers and $n$ nodes, each subcarrier is roughly shared by $\lceil  \frac{n}{m}  \rceil$ nodes. Starting from the beginning of the ordered subcarriers, each subcarrier is then assigned roughly to $\lceil  \frac{n}{m}  \rceil$ nodes that are not yet assigned a subcarrier starting from the beginning of the ordered nodes. In this way, we ensure that the widely available subcarriers are assigned to highly mobile nodes and the least widely available subcarriers are assigned to stationary or less mobile nodes.

\subsection{Handling Mobility across  SNOWs}
In this section, we present our approach to addressing the mobility across different SNOWs. Specifically, we  handle mobility problem that arises when a node goes out of the range of a BS. 
When a node goes out of the range of a BS, it needs to discover a new BS and get associated with it. Handoff becomes an issue when a node moves to an uncoordinated SNOW whose operating spectrum  is unknown.

For SNOW, the white space range is very wide, and the SNOW BS may be using a channel anywhere in that spectrum. A node operates on a narrowband subcarrier. Two subcarriers at center frequencies $f_i$ and $f_j$, $f_i\not=f_j$, are orthogonal when over time $T'$~\cite{ofdmbook}:
\begin{equation}\label{eqn:ortho}
\int_0^{T'}  \cos (2\pi f_i t) \cos (2\pi f_j t) dt=0.
\end{equation}

For example, when the overlap between subcarriers is 50\%, the BS bandwidth is 6 MHz, and the subcarrier width is 200 kHz, we can have 59 orthogonal subcarriers. Thus, to discover a new BS, it is very energy and time consuming for a low-power node as the node may need to scan thousands of subcarriers. Our approach has to deal with the following challenges as well. 
{\bf (1)} Spectrum dynamics due to primary user activity is handled using backup subcarriers in SNOW. However, such an approach does not work under mobility as the backup channels may be unavailable in a new location.  
A SNOW node has no access to the database and thus does not know the white space spectrum availability in its location. Spectrum sensing is highly energy consuming and is not feasible for it. {\bf (2)} It cannot transmit any probing message to explore a BS as it can interfere with primary users. The node hence needs to depend only on listening to SNOW's communication. {\bf (3)} The nearby BS may be using subcarriers of different bandwidth, and thus the node subcarrier may be unaligned (as depicted in Figure~\ref{fig:align}) and listening to nothing. Aligning with a BS channel is quite difficult as the BS subcarrier bandwidth is unknown to the moving node. {\bf (4)} The node should be able to distinguish between a primary user and a secondary user (BS). Our steps to address these challenges are as follows.

\subsubsection{BS Discovery}\label{subsubsec:discovery}
\begin{figure}[!htb]
\centering
\includegraphics[width=0.25\textwidth]{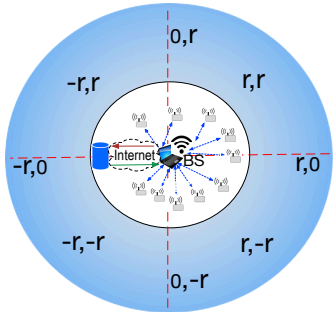}
\caption{Channel availability information in 8 locations.}
\label{fig:mobilityscenarioch}
\end{figure}

A direct approach to minimizing BS discovery overhead is that the current BS can provide a node, before it moves, the channels that the BS would find at 8 locations $(0, \pm r), (\pm r, 0), (\pm r, \pm r)$, considering its (estimated) communication range $r$ and location at $(0,0)$ assuming a Cartesian plane as shown in Fig.~\ref{fig:mobilityscenarioch}. After a node moves out of the current BS range, it can scan only those channels to find a neighboring BS. However, this approach would only work if it can inform the BS of its intention to move before it starts to move. Second, the node needs to know the direction of its movement and inform the BS.  Hence, we also propose another  energy-efficient and fast BS discovery technique that does not depend on these requirements. It utilizes the spectrum information by combining the received signal features to distinguish between primary users (TV stations) and a SNOW BS, and  considers the trade-off between discovery latency and energy consumption to allow efficient handoff management.

After a node goes out of range of its BS, it will scan one or more of its subcarriers. In either case, if it senses signal strength on a subcarrier, then it has to determine whether it is a BS or a primary user. If its subcarrier is not aligned with that subcarrier (in case of BS), it may not decode the received packets. To distinguish between the TV signal and the BS signal, we first need to detect the presence of primary users. The FCC regulation for protecting primary incumbents define a protection contour for TV station as the area where the received signal strength (RSS) is $ > -84$dBm~\cite{FCC_first_order}. We follow an approach similar to the one presented in Waldo~\cite{waldo}. Waldo's results show that low-cost sensors can efficiently detect white spaces ignored in the databases and existing approaches. Furthermore, depending on the white space device's antenna height, further separation (6 km for portable devices) is required to protect the primary incumbent. To detect white spaces, FCC recommends a typical antenna height of 10 meters. We consider an antenna height of 2 meters and compensate for the difference (8 meters) using the antenna correction factor using Hatas' urban area propagation model~\cite{waldo} considering $h_m$ as the antenna height in meters as follows.

\begin{equation}
a(h_m) = 3.2(\log 11.5 h_m)^2 - 4.97
\end{equation}

\begin{figure*}[!htb]
    \centering
      \subfigure[Channel detection before adding antenna correction factor\label{fig:tv2}]{
    \includegraphics[width=0.31\textwidth]{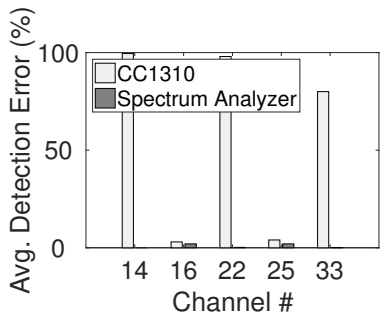}
      }
      \hfill
      \subfigure[Channel detection after adding antenna correction factor\label{fig:tv}]{
        \includegraphics[width=.31\textwidth]{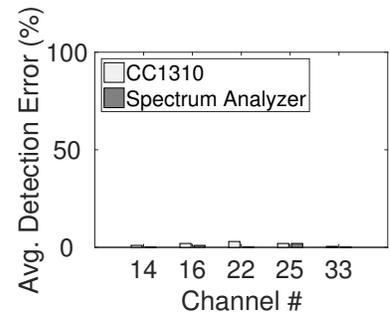}
      }
    \caption{Performance of TV detection}\vspace{-0.1in}
    \label{fig:whitespace}
\end{figure*}

Using Hata's model, the calculations result in $a(h_m) = 7.5$dB, which will be added uniformly to the RSS measurements. The addition of the antenna correction factor directly impacts the noisy measurements by making it closer to the threshold. Hence, improving the probability of false TV channel detection. We also follow Waldo's approach by considering the location safe for white space operation if the RSS $\geq -84$dBm, and the nearest measurement is 6 km away~\cite{waldo}. We record the measurements in a large metropolitan city for five different TV channels (14, 22, 33 are occupied by TV stations, 16 and 25 are white spaces). Additionally, we use the spectrum analyzer measurements and Google spectrum database as the ground truth to evaluate the SNOW nodes' TV channel detection performance. We collect 1500 spectrum measurements using four low-cost SNOW nodes (TI CC1310~\cite{cc1310}) over a period of 48 hours. Figure~\ref{fig:whitespace} shows the results for TV station detection. It is clear from Figure~\ref{fig:tv2} that without considering the antenna correction factor, CC1310 fails to detect TV transmission in all occupied channels. \revise{Operating in a low-frequency spectrum gives SNOW a tremendous obstacle penetration performance~\cite{snow}. Additionally, extensive experiments on TV detection performance is found in~\cite{waldo}.}

\begin{figure*}[!htb]
    \centering
      \subfigure[TV transmission\label{fig:tvtx}]{
    \includegraphics[width=0.35\textwidth]{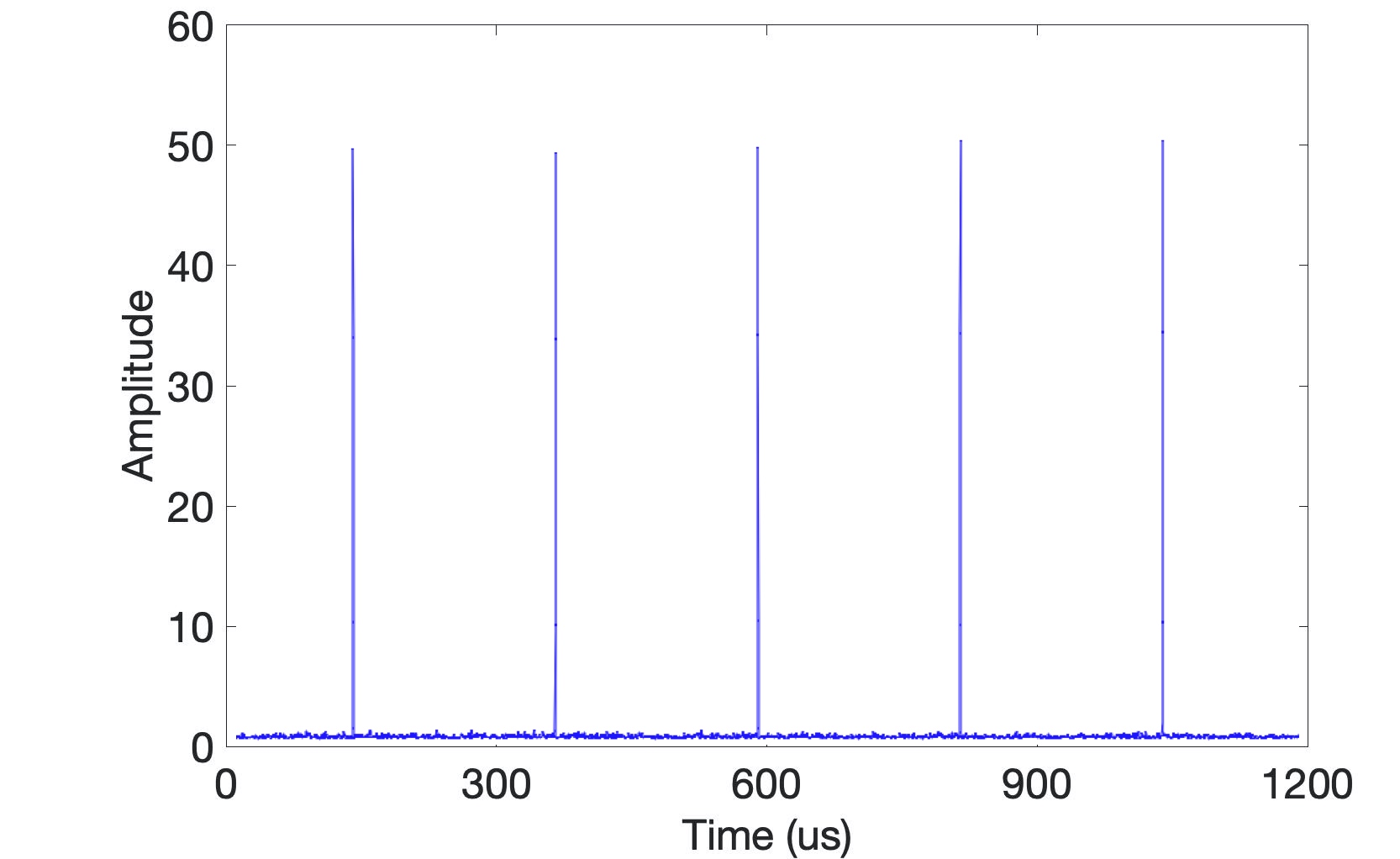}
      }
      \hfill
      \subfigure[SNOW transmission\label{fig:tvtx2}]{
        \includegraphics[width=.35\textwidth]{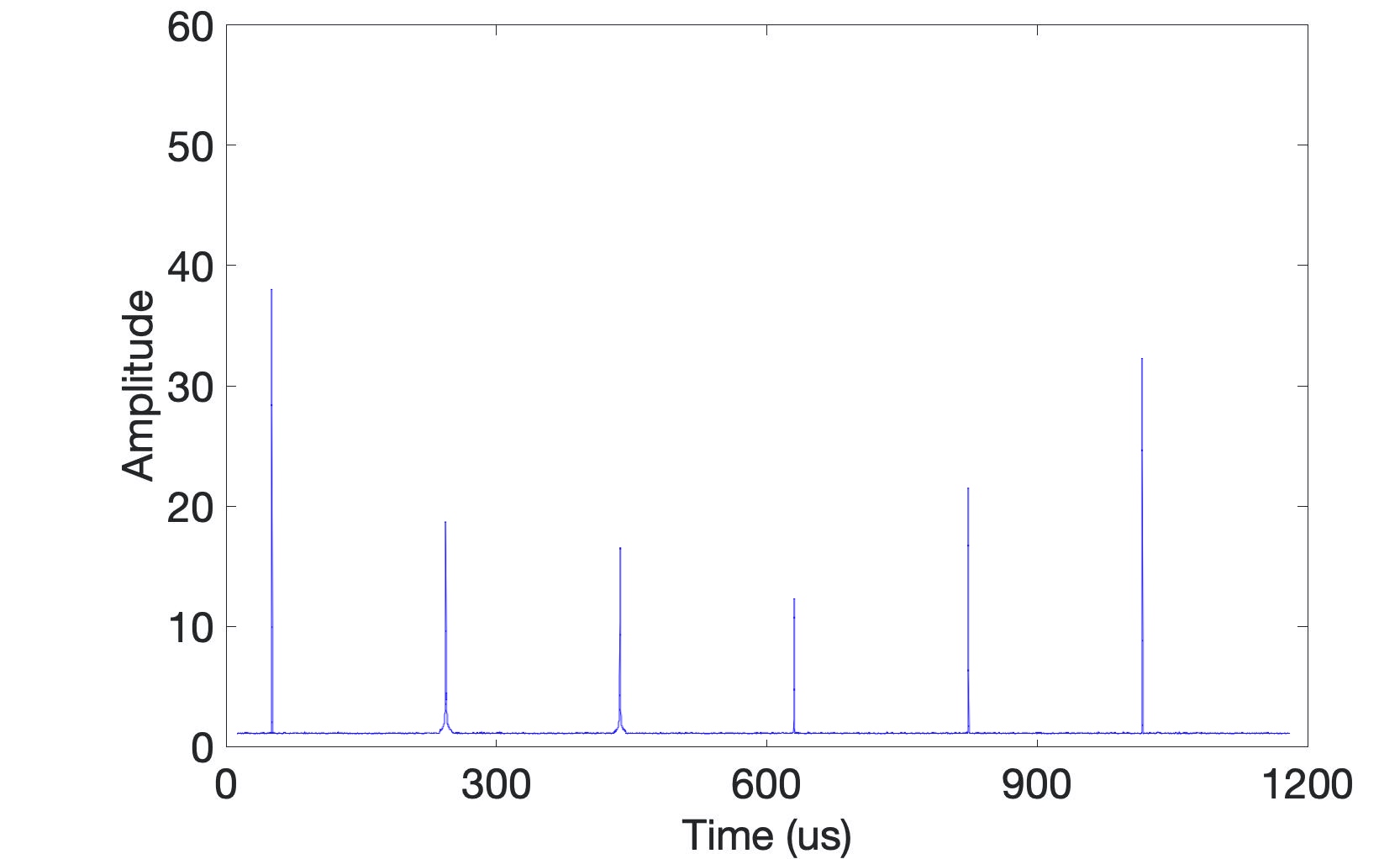}
      }
    \caption{TV and SNOW signal transmission recorded using a spectrum analyzer.}\vspace{-0.1in}
    \label{fig:tv_snow}
 \end{figure*}

We utilize a number of features of the received signals to distinguish between primary users (TV stations) and a SNOW BS. In addition to the common observation that RSS of the TV transmission is high and the signal amplitude is constant, primary user communication is observed to be continuous over a long duration (see Figure~\ref{fig:tvtx}). In contrast, SNOW BS signal amplitude is fluctuating during transmission and the BS may not have continuous communication for long periods as shown in Figure~\ref{fig:tvtx2}. In addition, if multiple consecutive channels have similar RSS, it is likely to be a BS because a BS typically uses more than one TV channel. For primary users, two consecutive channels should belong to two different primary users, and their signal strengths on two consecutive channels should be a lot different. To enable faster discovery, we also consider using a wider band for sensing, which will enhance the BS detection probability but will consume more energy. Since using a narrow subcarrier for searching can take a longer time, thus consuming much energy, such tradeoff is left as a design choice.

\subsubsection{Subcarrier Alignment} 
 Different SNOWs can have different subcarrier bandwidth, e.g., camera or audio may use a wider subcarrier. 
 Thus upon discovery of a new BS, as shown in Figure~\ref{fig:align}, a node's subcarrier may not be aligned with a BS subcarrier. Alignment is needed to start communication. Existing channel rendezvous techniques are not applicable as they consider the channels of equal bandwidth. Thus, this problem is specific to SNOW. By Equation (\ref{eqn:ortho}), an overlap can start from many points of a nearby subcarrier. Such an overlap makes the problem highly challenging. To solve the problem, we exploit several characteristics of SNOW design. Even though SNOW can use any subcarrier bandwidth, we consider that subcarrier bandwidth does not vary arbitrarily, and we assume each BS uses a subcarrier bandwidth from the values 100kHz, 200kHz, 400kH, or 600kHz. Upon discovering the presence of a BS, this assumption helps us simplify the synchronization with its subcarrier. This will be done using a wider bandwidth at the node and combining time and frequency domain energy-sensing.

The time-domain sensing is the typical carrier sensing that calculates the energy level using a moving average of the digital signals, i.e., the sequence of discretized, complex samples from the analog-to-digital converter, within a short period. A channel is considered {\slshape busy} if the output exceeds the predefined threshold.  The moving average's window size is set to half of the length of the preamble to ensure prompt sensing of a packet. Although time-domain sensing alone can sense a busy channel, it does not distinguish between different subcarriers. A node needs to analyze the frequency domain of the signals further. Specifically, it calculates the power spectrum density (PSD) of the recent $M$ samples using Fast Fourier Transformation (FFT). The node analyzes the power distribution and compares it with all possible channel-overlapping patterns based on the PSD. Intuitively, if the power is uniformly distributed over the entire spectrum, then the signals on the air come from a fully-overlapped subcarrier; otherwise, only a fraction of the channel is occupied. The exact fraction of channel in use is hard to calculate because different subcarriers may exhibit different power levels due to frequency-selective fading, and the imperfect hardware filter (used to confine the radio's bandwidth) spreads over the boundary of the PSD curve. But considering a limited number of bandwidths, the node can explore possible overlapping patterns and select the one with maximum matching with the PSD. The number of such patterns will also be limited as it is done after determining the presence of a BS.

\begin{figure}[!htb]
\centering
\includegraphics[width=0.33\textwidth]{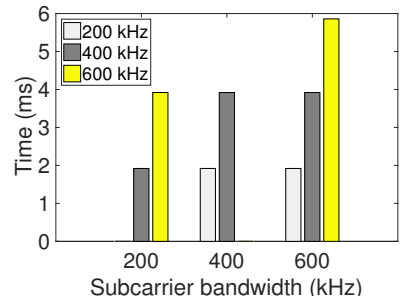}
\caption{Subcarrier alignment latency}
\label{fig:alignment}
\end{figure}

In SNOW, a node is less powerful and energy-constrained. The complexity of time-domain sensing is the same as the RSSI calculation in typical communications systems, which is linear with respect to the number of incoming samples. Since frequency sensing is performed only after a sequence of signals pass the time domain sensing, it takes constant time irrespective of the number of samples. The constant depends on the number of packets that cause the time-domain sensing to return {\slshape busy}. Note that such an approach is  needed only when a node moves to an uncoordinated SNOW. 
Once the node is aligned with any subcarrier of the new BS, it can use CSMA/CA approach to transmit to the BS and ultimately join the network. Figure~\ref{fig:alignment} shows the subcarrier alignment latency for different subcarrier bandwidths. 

\section{Experiments}\label{sec:experiments}

We have first implemented our mobility approaches using TI CC1310 devices as SNOW nodes. TI CC1310 is a tiny, low-cost, and low-power COTS device with a programmable PHY which was recently adopted as SNOW node \cite{iotdi2019}. \revise{To perform experiments at much longer communication ranges, we have also implemented our mobility approaches using USRP devices as SNOW nodes based on its current open-source implementation in GNU Radio \cite{opensource}.} GNU Radio is an open-source development toolkit that provides signal processing blocks to develop software-defined radio~\cite{gnuradio}. USRP is a hardware platform designed for RF application~\cite{usrp}. We have used two USRP 210 devices, each having a dual radio, as two BSs in each experiment for inter-SNOW mobility experiments. In the first set of experiments, we have used 10 TI CC1310 devices as SNOW nodes. In the other set of experiments, we have used 7 USRP 200 devices, each with a single radio, as SNOW nodes. The USRP devices operate in the band  70MHz -- 6GHz. Packets generation, decoder, and other implementation are adopted from SNOW open source implementation~\cite{opensource}.

Note that our experiments are performed mainly considering inter-SNOW mobility to show that our approach can enable such mobility. 
We cannot compare the results against the scenario when our approaches are not adopted because inter-SNOW mobility cannot be enabled without our approaches.  
However, we compare the performance against the stationary scenario to observe the performance degradation under mobility. In our experiments we shall demonstrate that such degradations are not high and our approaches show robustness in terms of reliability, latency, and energy consumption under various mobility scenarios.

\subsection{Default parameters}\label{subsec:default}
Parameters of interests are calibrated in different experiments based on requirements and the rest are left as defaults. The default experimental parameter settings are as follows.
\begin{itemize}
\item Frequency band: varying (470 MHz - 599 MHz)
\item Modulation: ASK/OOK
\item Packet size: 40 bytes
\item BS bandwidth: 6 MHz
\item Node bandwidth: TI CC1310: 200 kHz, USRP: 400 kHz
\item TX power: TI CC1310: 15 dBm, USRP: 0 dBm
\item Receiver sensitivity: -110 dBm
\item Distance: Indoor: 10 - 50 m, Outdoor: 900 m
\end{itemize} 

\subsection{Experiments with TI CC1310: Indoor and Outdoor Deployment}

\subsubsection{Indoor Deployment} 
The experiments with CC1310 were carried out in a hallway on the \revise{third floor inside the computer science building at Wayne State University.} We fixed the position of the BSs while a person is continuously moving at average walking speed from one end of the hallway to the other for 30 minutes. We kept the antenna height at 2 meters above the ground for all experiments. In all the experiments, the CFO and CSI are estimated and compensated based on SNOW implementation in~\cite{iotdi2019}. We used the default setting for all the experiments. 

\begin{figure}[!htb]
    \centering
      \subfigure[Reliability under mobility\label{fig:PER}]{
    \includegraphics[width=0.32\textwidth]{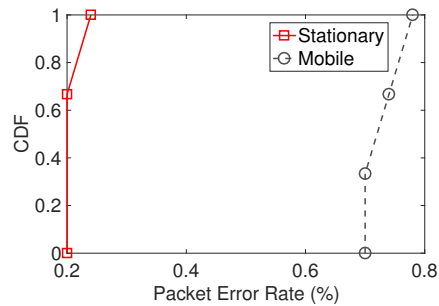}
      }
      \hfill
      \subfigure[Reliability with varying packet size\label{fig:psize}]{
        \includegraphics[width=.32\textwidth]{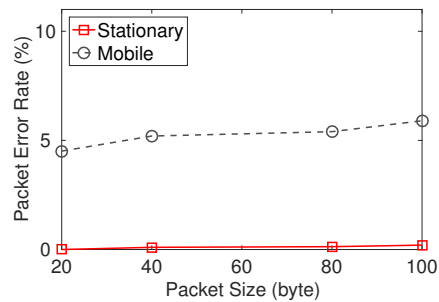}
      }
    \caption{Reliability under mobility and varying packet size.}
    \label{fig:reliability}
 \end{figure}

\noindent {\bf Reliability under Mobility.} 
We kept the distance between the node and BSs at approximately 10 meters to observe our proposed mobility approach's reliability. One node is stationary at this distance, and another node is continuously moving from one BS towards the other. The stationary node transmits 5000 packets to the BS while the mobile node transmits 2500 packets to the first BS and 2500 to the second BS after the joining process. The results in Figure~\ref{fig:PER} demonstrate that with minor human mobility, the Packet Error Rate (PER) slightly increases under our approach. For the stationary node, the PER is around 0.02\%, while the mobile node is 0.72\%. 
Also, we observe reliability with varying packet sizes. Figure~\ref{fig:psize} demonstrates the impact of packet size on our mobility approach. With 20-byte packet, the stationary node PER is 0\% (no packet loss). For the same packet size, the mobile node PER is around 4.5\%. 
Furthermore, for the 40-byte packet, the PER is 0.1\% and 5.2\% for the stationary and mobile node. With a 100-byte packet, the mobile node achieves 5.4\% PER, while the PER for the stationary node is 0.39\%. This result shows that packet size has an impact on reliability. Larger packets require more air time to receive, resulting in more interference leading to increased PER. \revise{The difference in the reliability performance between the two results is due to the channel condition changing.} For the mobile node, moving from one BS to another might increase the PER due to the channel condition at the new location, which might increase the PER. However, the results prove that our approach offers reliable communication under mobility. 

\begin{figure}[!htb]
	\centering
	\includegraphics[width=0.32\textwidth]{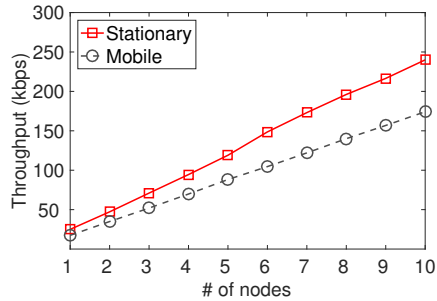}
	\caption{Throughput with varying \# of nodes.}\vspace{-0.1in}
	\label{fig:throughput}
\end{figure}

\noindent {\bf Maximum Achievable Throughput.}
The maximum achievable throughput is the total maximum number of bits the BS can receive per second. In this experiment, we calculate the maximum achievable throughput using our approach compared to the stationary nodes. In both scenarios, each node transmits 1000 40-byte packets. Figure~\ref{fig:throughput} shows that in a stationary scenario, the maximum achievable throughput is 240 kbps compared to 174 kbps during mobility when ten nodes transmit simultaneously. This result is not surprising since mobility increases the packet loss rate, which affects the throughput. 

\begin{figure}[!htb]
	\centering
	\includegraphics[width=0.32\textwidth]{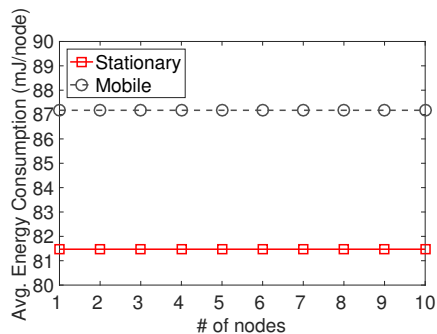}
	\caption{CC1310 Energy consumption}
	\label{fig:cceng}
\end{figure}

\begin{figure}[!htb]
	\centering
	\includegraphics[width=0.32\textwidth]{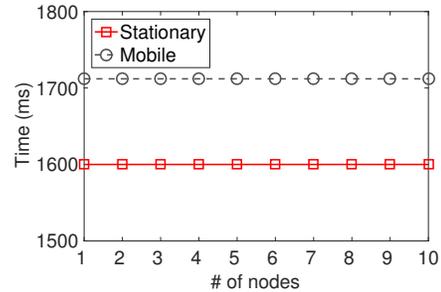}
	\caption{CC1310 Latency}
	\label{fig:cclat}
\end{figure}

\noindent {\bf Energy Consumption and Latency.}
To estimate the energy consumption and latency of CC1310 during mobility, we measure the average energy consumed at the nodes and the time it takes to transmit 1000 packets per node successfully. We placed ten nodes, each 50 m away from BS1. We performed two sets of experiments (stationary nodes and mobile nodes). In the mobility experiment, the nodes are placed 10 m away from BS1 and 50 m away from BS2. And the nodes move from BS1 to BS2. Also, we measure the overhead of the BS discovery and subcarrier alignment offline and add the results accordingly. To calculate the energy consumption, we use the energy model of CC1310 (Voltage is 3.8v, RX 5.4mA, and TX 13.4mA). We measure the time required to collect 10.000 packets at BS1 for the stationary nodes and 2500 packets and 7500 packets at BS1 and BS2, respectively, in the mobile scenario. Figure~\ref{fig:cceng} shows that in a stationary scenario, the average energy consumed by the node is 81.4mJ compared to 87.1mJ in during mobility. We observe similar behavior for the latency. As shown in Figure~\ref{fig:cclat}, the average latency for stationary nodes is 1.6s compared to 1.712s in mobile nodes. This result indicates that the number of nodes has an insignificant impact on the energy and latency regardless of mobility. This is due to the capability of SNOW BS, which allows multiple nodes to transmit in parallel. 
 
\subsubsection{Outdoor Deployment}
\revise{In this experiment, we evaluate our mobility approach's performance in the city of Detroit, Michigan (see Figure~\ref{fig:setup})}, in terms of maximum achievable throughput, energy consumption, and latency using CC1310 devices in outdoor deployments. We fix the location of the BSs and place the node inside a moving vehicle. The distance between the node and the BSs is approximately 900 meters. The vehicle speed varies between 5 mph and 40 mph. The data is collected at the BSs for further analysis.

\begin{figure}[!htb]
	\centering
	\includegraphics[width=0.32\textwidth]{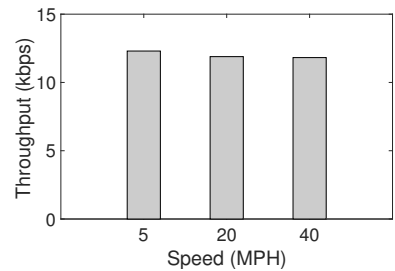}
	\caption{Throughput vs. node speed}\vspace{-0.1in}
	\label{fig:throughputout}
\end{figure}

\noindent {\bf Maximum Achievable Throughput.}  In this experiment, we compare the maximum achievable throughput at different speeds (5 mph, 20 mph, 40 mph). Each node transmits 1000 40-byte packets, and we calculate the combined throughput at the BSs. Figure~\ref{fig:throughputout} shows that the maximum achievable throughput is approximately 12.5 kbps at 5 mph speed compared to 11.89 kbps and 11.83 kbps At 20 mph and 40 mph, respectively. This shows that the speed (up to 40 mph) has an insignificant impact on the nodes' maximum achievable throughput.

\begin{figure}[!htb]
	\centering
	\includegraphics[width=0.32\textwidth]{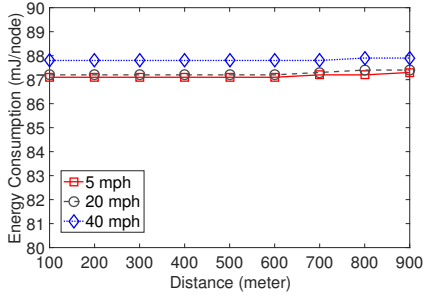}
	\caption{CC1310 Outdoor energy consumption}
	\label{fig:ccengout}
\end{figure}

\begin{figure}[!htb]
	\centering
	\includegraphics[width=0.32\textwidth]{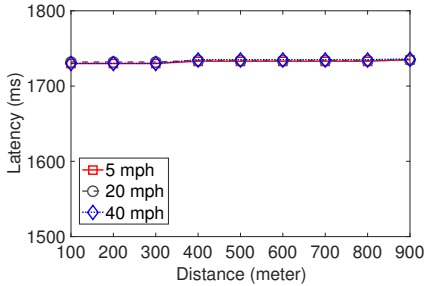}
	\caption{CC1310 Outdoor latency}
	\label{fig:cclatout}
\end{figure}

\noindent {\bf Energy Consumption and Latency.} To estimate the energy consumption and latency in outdoor deployment, we place the BSs at varying distances from the nodes (up to 900 meters). We set the nodes inside a moving vehicle. We measure the average energy consumed at each node and the time it takes to transmit 1000 packets per node successfully. Figure~\ref{fig:ccengout} shows that the average energy consumed by the node moving 5 mph is to 87.1mJ and 87.3 mJ at 100 m and 900 m, respectively. At 20 mph, the energy consumption is 87.1 mJ and 87.4 mJ at 100 m and 900 m, respectively. The average energy consumption is 87.8 mJ and 87.9 mJ at 100 m and 900 m for 40 mph. These results are similar to the indoor scenario where the node is moving at walking speed. In Figure~\ref{fig:cclatout}, the average latency for all nodes, regardless of the distance, is 1.735s. This result shows that the energy and latency for all nodes are similar except for the energy consumption at a 40 mph speed and a distance of 900 m where the results slightly vary due to higher PER.

\subsection{Experiments with USRP: Deployment in a Metropolitan City}
\begin{figure}[!htb]
	\centering
	\includegraphics[width=0.32\textwidth]{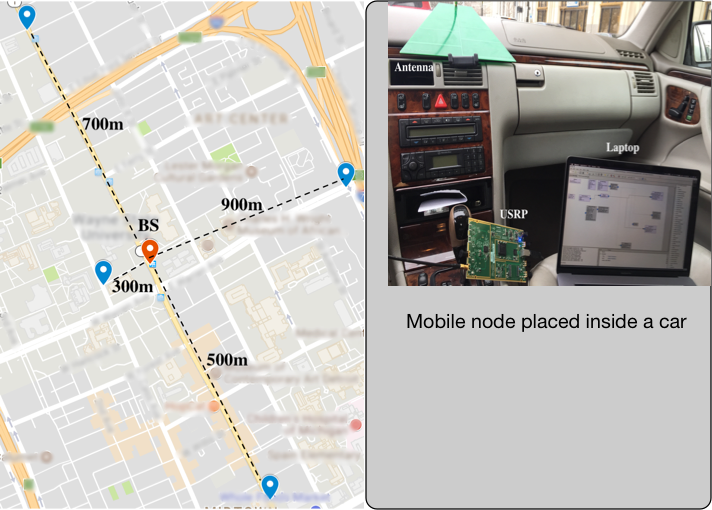}
	\caption{USRP experimental setup}
	\label{fig:setup}
\end{figure}

Figure~\ref{fig:setup} shows the distances in Detroit city in Michigan, where the mobility data were collected from nodes placed inside a moving car. The BSs are kept stationary. The vehicle is continuously moving at varying speeds (up to 40 mph) from one BS to the other in the mobile scenario. The antenna height was kept at 2 meters above the ground in all the experiments. We used the default setting in all the experiments. 

\subsubsection{Reliability over Distance}
\begin{figure}[!htb]
	\centering
	\includegraphics[width=0.32\textwidth]{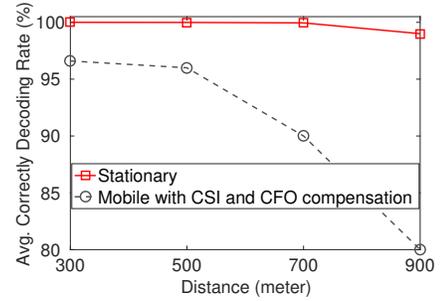}
	\caption{Reliability over distances}
	\label{fig:distance}
\end{figure}

To observe the effect of distance on the reliability of SNOW in mobile scenarios, we collect the data at 300m, 500m, 700m, and 900m from the BS, respectively. Each node transmits 5000 packets. To measure the reliability, we chose Correctly Decoding Rate (CDR), which is the ratio of the number of correctly decoded packets at the BS to the total number of transmitted packets~\cite{snow2}. Figure~\ref{fig:distance} shows the reliability over various distances from the BS when the node is moving from one BS to the other. At 300 meters, the BSs can decode on average 96.6\% of the packets from the mobile node compared to 100\% for the stationary node. Furthermore, at 500 meters away, the mobile node's reliability reduces to 96\%, while the stationary node achieves 99.99\% reliability. At 900 meters, the reliability is 80\% for the mobile node compared to 99.95\% at the stationary node. These results show that the distance between the mobile node and BS has a significant impact on decoding reliability. However,   Even for the stationary node, its' performance is slightly impacted by the distance from the BS.   

\subsubsection{Performance of SNOW with CFO}
\begin{figure}[!htb]
	\centering
	\includegraphics[width=0.32\textwidth]{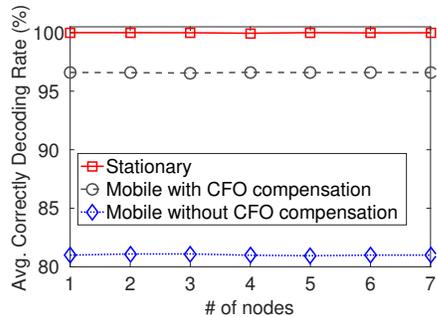}
	\caption{Performance under mobility with CFO}
	\label{fig:cfo2}
\end{figure}

In this experiment, we observe the performance of SNOW to demonstrate the effect of CFO estimation and compensation in mobile environments. We compare the CDR of a mobile SNOW node in two cases, with CFO estimation and compensation and without CFO compensation. Also, we compare the performance of each case to the performance of a stationary SNOW node. All the nodes were 500m away from the BS. The mobile nodes were placed in a car moving at varying speeds. Each node transmits 5000 packets asynchronously to the BSs. For mobile nodes, each node transmits 2500 packets to BS1 and 2500 to BS2. Figure~\ref{fig:cfo2} demonstrates the effect of CFO under mobility. For stationary nodes, the average CDR is around 99.97\% for all the transmitted packets. Without compensation for CFO, the average CDR is around 80\% for all the nodes. However, we compensate for CFO; the average CDR increases to 96\%, which is significant. This result demonstrates that in mobile environments, CFO could severely impact the transmission reliability. Thus, CFO estimation and compensation could significantly increase the reliability of the transmission in inter-SNOW communication. 

\subsubsection{Maximum Achievable Throughput}
In this experiment, we compare the maximum achievable throughput in mobile inter-SNOW with the stationary SNOW. For both scenarios, each node transmits 100 40-byte packets. We calculate the combined throughput at the BSs. Figure~\ref{fig:throughput} shows that when 8 nodes are transmitting, the maximum achievable throughput is 298 kbps and 393 kbps for mobile and stationary SNOWs, respectively. During mobility when 10 nodes transmit simultaneously. Due to the increased packet loss rate during mobility, stationary SNOW achieves better throughput. 
\begin{figure}[!htb]
	\centering
	\includegraphics[width=0.32\textwidth]{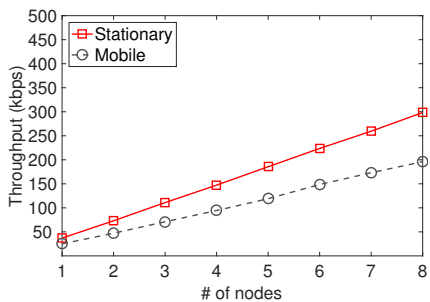}
	\caption{Throughput vs \# of node}
	\label{fig:tp}
\end{figure}

\subsubsection{Energy Consumption and Latency}
In this experiment, we demonstrate the efficiency of our mobility approach for USRP in terms of energy consumption and latency. Specifically, we compare the efficiency of mobile SNOW with stationary SNOW. We observed that the performance of SNOW under mobility is affected by the distance from the BS. Hence, for a fair comparison with stationary SNOW, we place 7 mobile node 900m away from the BS2 while continuously moving at approximately 20mph towards BS1. Furthermore, since USRP devices allow for bidirectional communication, each node transmits 100 packets (50 to BS1 and 50 to BS2 during mobility) during the upward duration (1s) and waits until the end of the upward duration to receive an acknowledgment (ACK) from the BSs. We then calculate the average energy consumption per node and the time needed to collect all the packets at the BS. 

\begin{figure}[!htb]
	\centering
	\includegraphics[width=0.32\textwidth]{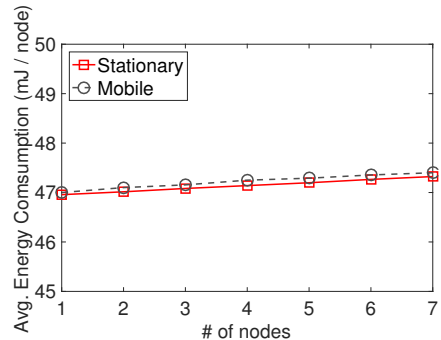}
	\caption{Energy consumption}
	\label{fig:eng}
\end{figure}

\begin{figure}[!htb]
	\centering
	\includegraphics[width=0.32\textwidth]{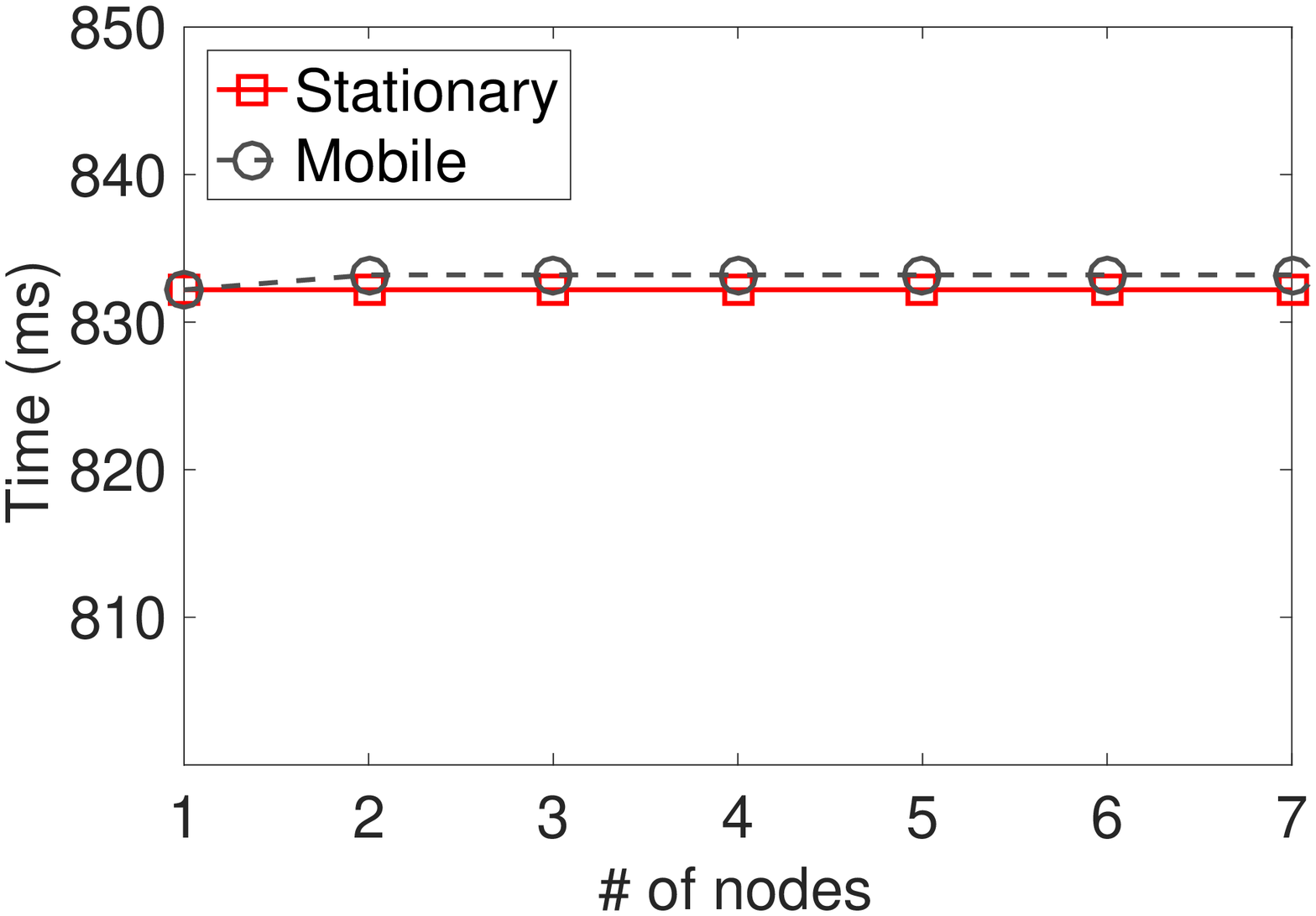}
	\caption{Latency}
	\label{fig:lat}
\end{figure}

Figure~\ref{fig:eng} shows that the average energy consumed at the mobile nodes is around 47.4mJ compared to 47.32\% in stationary nodes when 7 nodes transmit. This shows that mobility has a minimal impact on the energy efficiency of the node. Similar to the average energy consumption, Figure~\ref{fig:lat} shows that the latency of collecting all packets in mobile SNOW is comparable to the stationary SNOW. These results demonstrate that the efficiency of SNOW is not affected by mobility.   

\section{Conclusions}\label{sec:conclude}

 In this paper, we have addressed mobility in SNOW (Sensor Network Over White spaces), an LPWAN that is designed based on D-OFDM and that operates in the TV white spaces. 
SNOW supports massive concurrent communication between a base station (BS)  and numerous nodes.  We have proposed a dynamic CFO estimation and compensation technique 
to handle mobility impacts on ICI.  We have also proposed to circumvent the mobility impacts on geospatial variation of white space through a mobility-aware spectrum assignment to nodes.  To enable mobility  of the nodes across different SNOWs, we have proposed an efficient handoff management through a fast and energy-efficient BS discovery and quick association with the BS by combining time and frequency domain energy-sensing. Experiments through SNOW deployments in a large metropolitan city and indoors have shown that our proposed approaches  enable mobility across multiple different SNOWs and provide robustness in terms of reliability, latency, and energy consumption under mobility.

%
%
\section{Acknowledgments}
This work was supported by NSF through grants CAREER-1846126, CNS-2006467, and by Wayne State University through the Rumble Fellowship.

%
%
\balance
\bibliographystyle{abbrv}
\bibliography{ewsn_bib,whitespacebib}  
\end{document}